\documentclass[prl,11pt,superscriptaddress,floatfix,nofootinbib,notitlepage,longbibliography]{revtex4-1}

\usepackage{graphicx}
\usepackage[dvipsnames]{xcolor}
\usepackage[colorlinks=true,allcolors=RoyalBlue]{hyperref}
\usepackage{amsmath}
\usepackage{amssymb}
\usepackage{soul}

\begin{document}

\title{Atomically sharp 1D interfaces in 2D lateral heterostructures \\of VSe$_2$---NbSe$_2$ monolayers}
\author{Xin Huang}
\thanks{These two authors contributed equally}

\affiliation{Department of Applied Physics, Aalto University, FI-00076 Aalto, Finland}

\author{H\'ector Gonz\'alez-Herrero}
\thanks{These two authors contributed equally}

\affiliation{Department of Applied Physics, Aalto University, FI-00076 Aalto, Finland}
\affiliation{Departamento F\'isica de la Materia Condensada, Universidad Aut\'onoma de Madrid, Madrid E-28049, Spain}

\author{Orlando J. Silveira}
\affiliation{Department of Applied Physics, Aalto University, FI-00076 Aalto, Finland}

\author{Shawulienu Kezilebieke}
\affiliation{Department of Physics, Department of Chemistry and Nanoscience Center, 
University of Jyväskyl\"a, FI-40014 University of Jyväskyl\"a, Finland}

\author{Peter Liljeroth}
\affiliation{Department of Applied Physics, Aalto University, FI-00076 Aalto, Finland}

\author{Jani Sainio}
\affiliation{Department of Applied Physics, Aalto University, FI-00076 Aalto, Finland}

\begin{abstract}
Van der Waals heterostructures have emerged as an ideal platform for creating engineered artificial electronic states. While vertical heterostructures have been extensively studied, realizing high-quality lateral heterostructures with atomically sharp interfaces remains a major experimental challenge. Here, we advance a one-pot two-step molecular beam lateral epitaxy approach and successfully synthesize atomically well-defined 1T-VSe$_2$---1H-NbSe$_2$ lateral heterostructures. We demonstrate the formation of defect-free lateral heterostructures and characterize their electronic structure using scanning tunnelling microscopy and spectroscopy together with density functional theory calculations. We find additional electronic states at the one-dimensional interface as well as signatures of Kondo resonances in a side-coupled geometry. Our experiments explore the full potential of lateral heterostructures for realizing exotic electronic states in low-dimensional systems for further studies of artificial designer quantum materials.
\end{abstract}
\maketitle

\section{Introduction}
Heterostructures of two-dimensional (2D) materials are seen as one of the most flexible platforms to study correlated electronic states and realize novel phenomena in condensed matter systems \cite{Geim2013,Novoselov2016,Castellanos-Gomez2022}. Most van der Waals (vdW) heterostructures are assembled through vertical stacking, where layers interact only via van der Waals forces. These vertical heterostructures \textit{de facto} realize an effective 2D system. In addition to these 2D systems, it would be desirable to have access to one-dimensional (1D) structures, where different electronic phenomena can arise. 1D lattices with a lower dimensional structure also provide a simpler prototype to understand many-body physics. However, dimensionality reduction starting from a higher dimension remains challenging; it is inherently difficult to fabricate 1D structures with top-down methods, \textit{e.g.~}mechanical exfoliation and transfer. Currently, almost all experimentally realized 1D structures in 2D materials are naturally occurring, for example grain boundaries or domain walls \cite{Barja2016,Yin2016,Cho2017,Yuan2018}.

On the other hand, fabricating lateral heterostructures by bottom-up synthesis offers intriguing possibilities to create 1D structures. Compared to their vertical counterparts, lateral heterostructures or in-plane heterojunctions have covalent bonds between the components, and can form artificial 1D structures capable of hosting novel electronic states. However, the most common method to produce lateral heterostructures---chemical vapour deposition (CVD)---has considerable drawbacks: the atomic-scale structure of CVD grown interfaces typically suffers from a high density of imperfections, such as elemental doping and alloying, various defects and dislocations \cite{Duan2014,Huang2014,Gong2014,Lin2014,Li2015,Zhang2017,Sahoo2018,Zhang2018,Lin2018,Wang2019,Zhang2021,Avalos-Ovando2019}. In addition, most attention on lateral heterostructures in transition metal dichalcogenides (TMDCs) to date has focussed on homophase semiconductor-semiconductor junctions (\textit{e.g.} MoS$_2$, MoSe$_2$, WS$_2$, WSe$_2$), where the two components have the same crystal structure, \textit{e.g.} 1H- with 1H-phase. \cite{Duan2014,Huang2014,Gong2014,Lin2014,Chen2015,Li2015,Zhang2017,Sahoo2018,Zhang2018,Lin2018}

In this work, we choose two heterophase metallic TMDC monolayers, vanadium diselenide (VSe$_2$) and niobium diselenide (NbSe$_2$), with different crystal phases of 1T and 1H with both having electronic structures where electron correlations play a significant role.
VSe$_2$ is metallic in its monolayer octahedral 1T structure, and it has been reported to have various magnetic ground states competing with charge density wave (CDW) order depending on factors such as defect density, doping and strain \cite{Feng2018,Coelho2019,Wong2019,Fumega2019,Fumega2023,Ma2012,Shawulienu2020,Chua2020,Memarzadeh_2021}. The other ingredient, 1H-phase of NbSe$_2$, is a 2D metal exhibiting significant electron correlations, and CDW and superconducting orders at low temperatures (superconducting $T_c \sim1$ K for a monolayer) \cite{Ugeda2015,Xi2015,Zhao2019,Divilov2021,Ganguli2022,Wan2022,Akber2024}.

Here, we demonstrate a one-pot two-step lateral epitaxy technique to fabricate atomically sharp and well-defined lateral heterostructures of 1T-VSe$_2$---1H-NbSe$_2$ by molecular beam epitaxy (MBE). We probe them by low-temperature scanning tunneling microscopy (STM) and spectroscopy (STS) and identify two different 1D interface structures corroborated by density-functional theory (DFT) calculations. These heterostructures exhibit 1D interfacial states and in addition signatures of Kondo resonances which to our knowledge is the first such observation in an atomic-scale side-coupled geometry. This work demonstrates a new approach for achieving complex lateral heterostructures with atomically well-defined 1D interfaces where it is possible to realize correlated many-body states via lateral coupling.

\section{Synthesis and structure of the lateral heterostructures}

\begin{figure}[t!]
\centering
\includegraphics[width=1\textwidth]{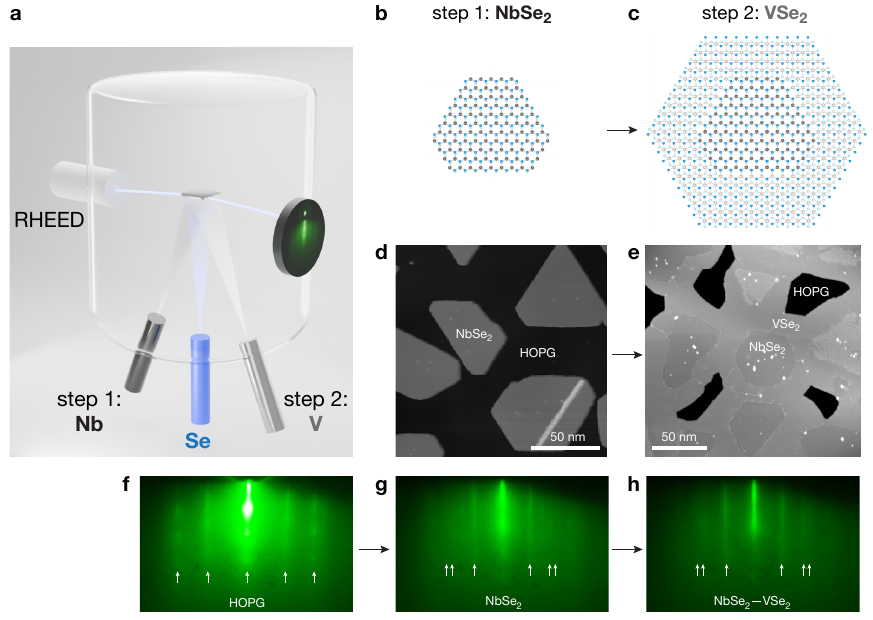}
\caption{\label{fig:fig1}Synthesis of the lateral heterostructures. \textbf{a}, Illustration of our one-pot two-step lateral epitaxy. \textbf{b}, Schematic of a NbSe$_2$ island, and \textbf{c}, in-registry lateral epitaxy of VSe$_2$ surrounding the NbSe$_2$ island. \textbf{d}, Scanning tunnelling microscopy (STM) topography image of the sample after step 1: growth of NbSe$_2$. $V_\text{s}$=+1 V, $I_\text{t}$= 2 pA. \textbf{e}, STM image after step 2: growth of VSe$_2$. $V_\text{s}$=$-$1.5 V, $I_\text{t}$=10 pA. \textbf{f}--\textbf{h}, Reflection high-energy electron diffraction (RHEED) patterns of the substrate HOPG (f), monolayer NbSe$_2$ islands (step 1, g), and VSe$_2$—NbSe$_2$ lateral heterostructures (step 2, h).
}
\end{figure}

We overcome the challenges in CVD \cite{Duan2014,Huang2014,Gong2014,Lin2014,Chen2015,Li2015,Zhang2017,Sahoo2018,Zhang2018,Lin2018} by introducing a one-pot two-step lateral epitaxy method utilising MBE (Fig.~\ref{fig:fig1} and Methods). The first step is to grow monolayer islands of 1H-phase NbSe$_2$ at 550 °C with well-defined and straight edges (Fig.~\ref{fig:fig1}b, d). The second step is the lateral epitaxy growth of 1T-VSe$_2$ at 375 °C (Fig.~\ref{fig:fig1}c, e and Supplementary Fig.~S2), using the 1H-NbSe$_2$ islands' edges as seeds \cite{Liu2014,Sutter2014}. Growth temperature and growth sequence play a crucial role \cite{Zhou2023} in lateral epitaxy:
the material with higher growth temperature should be synthesized first in order to preserve island morphology, the edge structure, and to prevent unwanted alloying in the next steps.
Our molecular beam lateral epitaxy growth can be monitored \textit{in situ} via reflection high-energy electron diffraction (RHEED). During step 1, the RHEED pattern gradually develops stripes of NbSe$_2$ in addition to the pattern related to the highly oriented pyrolytic graphite (HOPG) substrate (Fig.~\ref{fig:fig1}f, g). Subsequently, during step 2, the RHEED pattern of the newly formed lateral heterostructure basically overlaps with the one of the NbSe$_2$, because the nominal lattice constant of VSe$_2$ is very similar to that of NbSe$_2$ ($\sim$0.1 Å difference) and the overall increasing coverage of the monolayer leads to dimming and the eventual disappearance of the HOPG RHEED pattern (Fig.~\ref{fig:fig1}h and Supplementary Fig.~S1).

\begin{figure}[h!]
\centering
\includegraphics[width=1\textwidth]{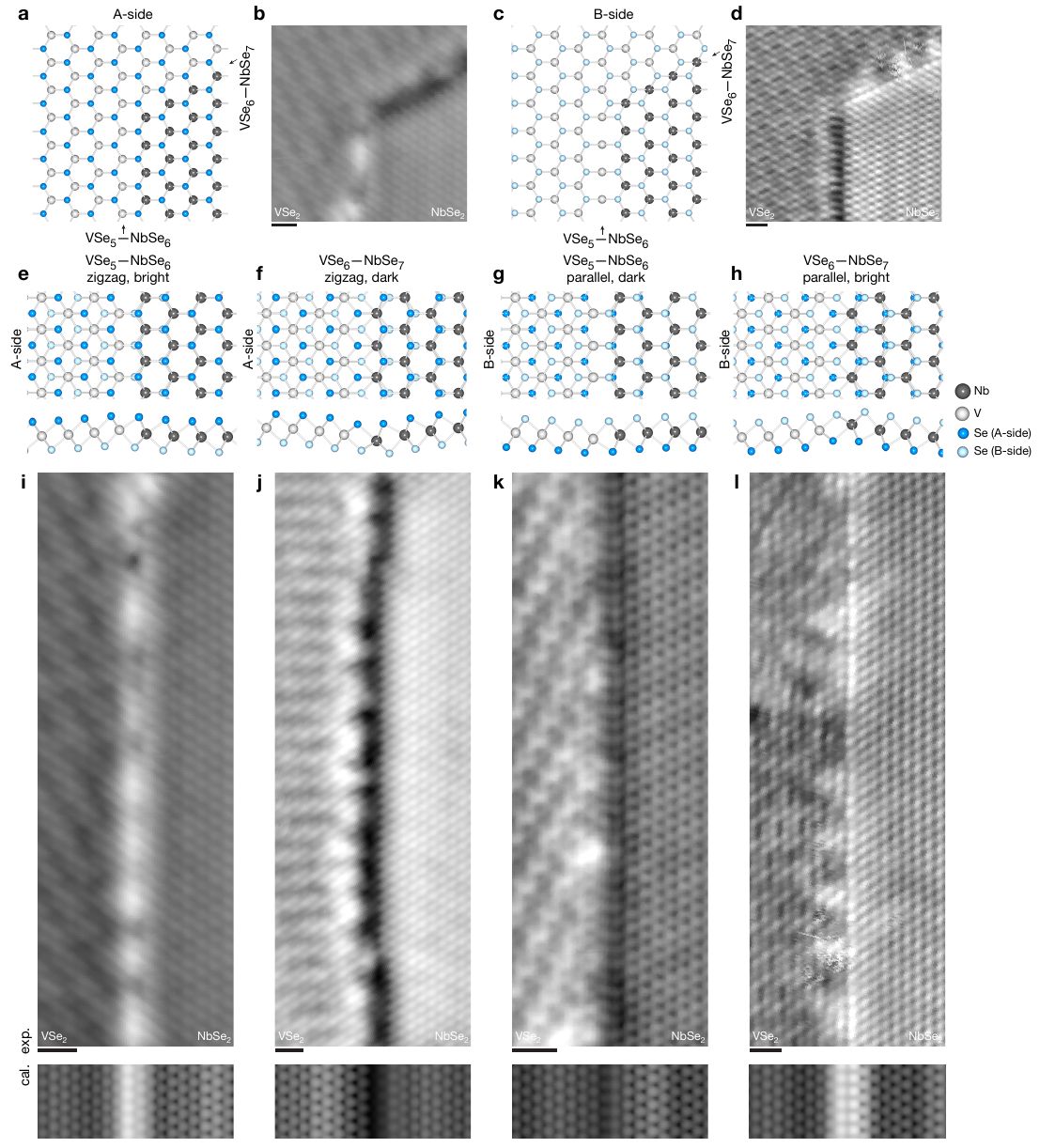}
\caption{\label{fig:fig2} Two types of lateral heterostructures. 
\textbf{a}, A-side view (with A-side Se atoms only) and \textbf{b}, its STM topography image. 
\textbf{c}, B-side view (with B-side Se atoms only) and \textbf{d}, its STM topography image. 
\textbf{e--h}, Top and side view of the two types of lateral heterostructures, 
and \textbf{i--l}, their corresponding atomically resolved STM topography images (experimental and calculated). 
Scan parameters: 
\textbf{b}, \textbf{d}, \textbf{i}, \textbf{j}, \textbf{l}: $V_\text{s}$=$-$1 V, $I_\text{t}$= 100 pA. \textbf{k}: $V_\text{s}$=$-$0.99 V, $I_\text{t}$= 50 pA.
Calculated STM images, $V_\text{s}$=$-$0.5 V.
All scale bars are 1 nm.} 
\vspace{18pt}
\end{figure}

We first confirm the growth of 1H-NbSe$_2$ and 1T-VSe$_2$ by their STM topography images with typical charge density waves (3$\times$3 for 1H-NbSe$_2$ \cite{Ugeda2015}, $\sqrt{3}\times2$ and $\sqrt{3}\times\sqrt{7}$ for 1T-VSe$_2$ \cite{Chen2018,Coelho2019,Chua2020}. See Supplementary Fig.~S2). Creating lateral heterostructures from such systems provides new possibilities for research on 2D charge density wave orders. In our heterostructures, the intrinsic CDWs of both materials extend right up to the interface (Fig.~\ref{fig:fig2} and Supplementary Fig.~S3--S7 and Fig.~S10), where they abruptly switch from the characteristic CDW of 1H-NbSe$_2$ to that of 1T-VSe$_2$ in contrast to a similar system showing a CDW proximity effect \cite{Akber2024}. The undeformed CDWs indicate that there is no observable in-plane lattice distortion, implying lack of significant amounts of e.g.~strain or doping being induced at the interface, as especially the CDW in 1T-VSe$_2$ should be sensitive to that \cite{Fumega2023}.

We find two different types of lateral heterostructures as shown in Fig.~\ref{fig:fig2}. The 1D interfaces are atomically sharp without cross-contamination, doping, or alloying, with lengths extending up to $\sim$ 20--40 nm (Fig.~\ref{fig:fig2} and Supplementary Fig.~S3--S5), which to our knowledge are the longest high-quality TMDC lateral heterostructures grown so far \cite{Avalos-Ovando2019}. The first-grown 1H-NbSe$_2$ islands mostly have a hexagonal shape with 120° corners. Considering the crystal structure, this means that the neighboring edges are crystallographically distinct, \textit{i.e.}~with alternating edge terminations with \textit{e.g.} Nb- or Se-terminated edges \cite{Lu2017,Zhang2022}. The subsequent epitaxy of 1T-VSe$_2$ will most likely form Nb-Se-V chemical bonds over the interface, especially in a Se-rich growth environment.
Combining this with the 1T-VSe$_2$ crystal structure, we can identify four different types of possible lateral heterostructures, with vanadium and niobium coordination numbers between 5 and 7 (Supplementary Fig.~S13 and Fig.S15). Corroborated by our DFT calculations we indentify the two most stable heterostructures, and we label them as VSe$_5$---NbSe$_6$ and VSe$_6$---NbSe$_7$ (Fig.~\ref{fig:fig2}a, c). Also other type of structures have been considered, but they fail in either producing the correct type of atomic arrangement seen in Fig.~\ref{fig:fig2} or they are inconsistent with in-registry growth around 120° corners of 1H-NbSe$_2$ islands (Supplementary Fig.~S13). 

However, depending on which side of these structures grows on the substrate (and hence which side faces the STM tip), these two types of lateral heterostructures show four different STM topographies (arising mostly from the top layer of Se atoms), as shown in Fig.~\ref{fig:fig2}e--h and labelled zigzag or parallel with bright or dark contrast.
Hence, the two different interface structures for VSe$_5$---NbSe$_6$ and VSe$_6$---NbSe$_7$ can appear  originating from two different sides of the monolayer, labelled A-side and B-side. 
If imaged on the A-side, the VSe$_5$---NbSe$_6$ interface shows a bright zigzag structure, and VSe$_6$---NbSe$_7$ a dark zigzag; if scanning on the B-side, VSe$_5$---NbSe$_6$ shows a dark parallel structure, and VSe$_6$---NbSe$_7$ a bright parallel one.

Our DFT calculations also give the correct contrast from the top layer Se atoms in simulated STM images compared to experimental ones (Fig.~\ref{fig:fig2}i--l).
The DFT calculations suggest that the interfaces have a slight structural deformation producing a structure similar to the 1T$'$-phase, and the trend of bright/dark contrast comes in part from a small corrugation at the interface (Fig.~\ref{fig:fig2}e--h). The contrast also has an electronic component discussed later. We find that the bright and dark contrasts alternate between the adjacent edges of hexagonal 1H-NbSe$_2$ islands (see Fig.~\ref{fig:fig2}b, d, and Supplementary Fig.S3--S6, and Fig.S9). In addition, interfaces around a single 1H-NbSe$_2$ island are either all zigzag or all parallel. These two experimental observations further support the structural assignment above and are consistent with our DFT calculations (Fig.~\ref{fig:fig2}a, c).

\section{Electronic structure of the 1D interfaces}
\begin{figure}[h!]
\centering
\includegraphics[width=1\textwidth]{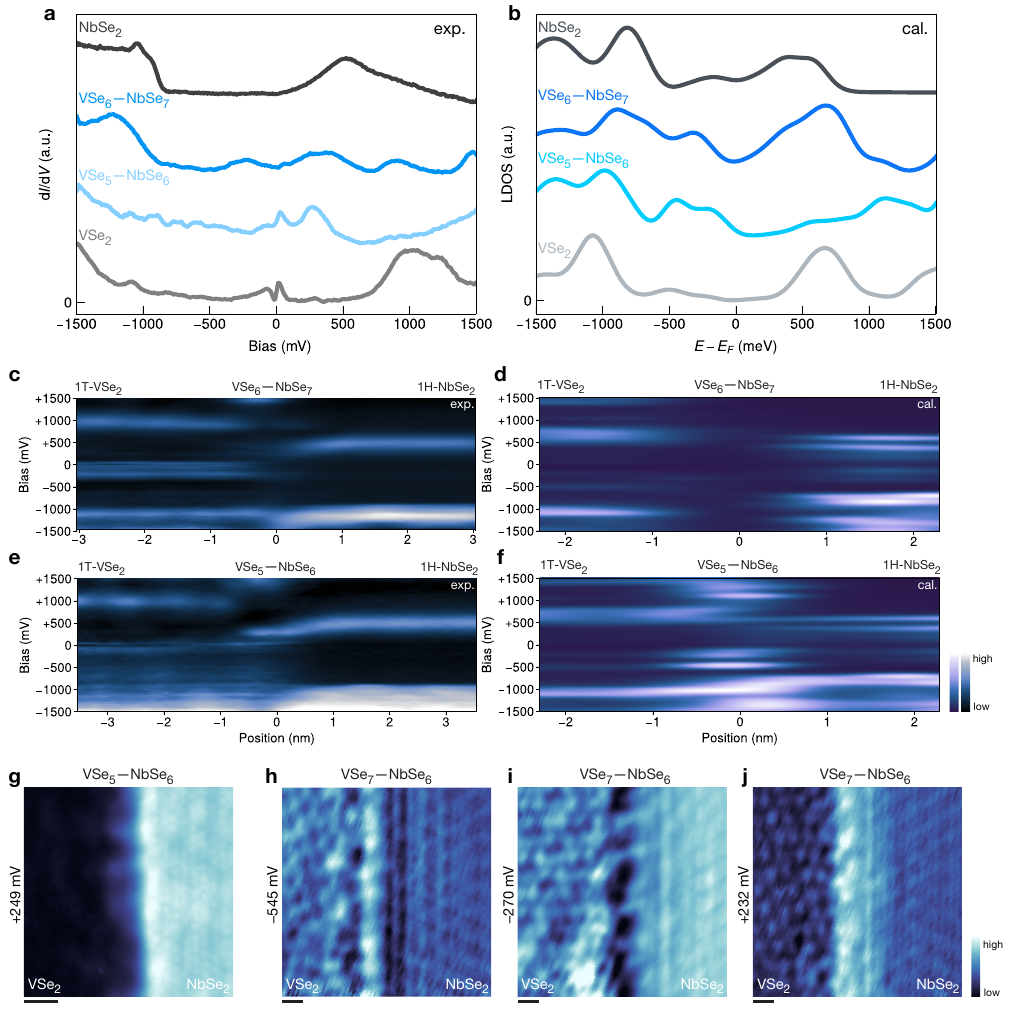}
\caption{\label{fig:fig3} Electronic structure of the lateral heterostructures. \textbf{a}, d$I$/d$V$ spectra of 1H-NbSe$_2$, 1T-VSe2$_2$ and VSe$_5$---NbSe$_6$ and VSe$_6$---NbSe$_7$ interfaces (shifted vertically for clarity). \textbf{b}, DFT calculated LDOS of 1H-NbSe$_2$, 1T-VSe2$_2$ and VSe$_5$---NbSe$_6$ and VSe$_6$---NbSe$_7$ interfaces (shifted vertically for clarity). \textbf{c-f}, experimental d$I$/d$V$ spectra and calculated LDOS along a line across the VSe$_6$---NbSe$_7$ (panels c and d) and VSe$_5$---NbSe$_6$ (panels e and f) interfaces. \textbf{g}, d$I$/d$V$ map of the 1D interfacial state of VSe$_5$---NbSe$_6$, at +249 mV ($V_\text{mod}$=4 mV).
\textbf{h}--\textbf{j}, constant-current d$I$/d$V$ maps of VSe$_6$---NbSe$_7$ interface at $-$545 mV, $-$270 mV and +232 mV respectively ($V_\text{mod}$=10 mV), which show additional LDOS modulation inside 1H-NbSe$_2$ in the vicinity of the VSe$_6$---NbSe$_7$ interface. All scale bars are 1 nm.}
\vspace{18pt}
\end{figure}

We focus next on the electronic behaviour of the two lateral heterostructures on the A-side with zigzag interfaces (Fig.~\ref{fig:fig3}). The differential conductance (d$I$/d$V$) spectrum of monolayer 1H-NbSe$_{2}$ shows a peak at around +500 mV related to the Nb-based conduction band, and the peak $\sim$ $-$1100 mV is related to its valence band \cite{Ugeda2015,Silva-Guillen2016}. The conduction band crosses the Fermi level away from the $\Gamma$-point and it is difficult to resolve the bottom of the conduction band \cite{Ugeda2015}. Together with the gap between the bottom of the conduction band and the valence band, this results in a bias region with low d$I$/d$V$ signal (from $\sim$ 0 mV to $\sim$ $-$840 mV).
The d$I/$d$V$ of 1T-VSe$_{2}$ is consistent with the results reported in the literature \cite{Pasztor2017,Jolie2019_VSe2,Wong2019,Shawulienu2020}, with the signal close to the Fermi level arising from the vanadium $d$-states and V-shaped features very close to $E_\mathrm{F}$ resulting from the CDW \cite{Jolie2019_VSe2,Wong2019}.

The two interfaces show quite different features compared to each other and the corresponding 1H-NbSe$_2$ and 1T-VSe$_2$ bulk monolayers. For the VSe$_5$---NbSe$_6$ interface, a prominent peak shows up around +250 mV (Fig.~\ref{fig:fig3}a). From the differential conductance (d$I$/d$V$) map taken at this energy (Fig.~\ref{fig:fig3}g) we can confirm this state is localized at the interface, with a spatial extent of roughly 1 nm. For VSe$_6$---NbSe$_7$, no new states are observed but the Nb-based conduction band shifts to lower energy very close to the interface while the valence band of 1T-VSe$_2$ remains at the same energy. The evolution of the electronic states can be also visualized by recording spectra along a line across the interface, shown in Figs.~\ref{fig:fig3}c, e.  

DFT calculated LDOS spectra (Fig.~\ref{fig:fig3}b, d and f, and Supplementary Fig.~S12) show similar behaviour. The additional electronic state of VSe$_5$---NbSe$_6$ is reproduced in the calculated LDOS (with an energy shift which is also seen for the monolayer 1T-VSe$_2$ states). This state is absent on the VSe$_6$---NbSe$_7$ interface.

We also calculated the projected density of states (PDOS) on the top layer Se atoms at the VSe$_5$---NbSe$_6$ interface (see Supplementary Fig.~S14), since they contribute most to the STM signal. The calculations indicate that the bright contrast at the interface is not only due to the corrugation of the Se atoms but also to changes in the electronic structure. The same is true for the VSe$_6$---NbSe$_7$ interface (dark contrast) but in that case the effect of the corrugation appears to be larger.

Finally, we observed additional density of states oscillations parallel to the VSe$_6$---NbSe$_7$ interface on the 1H-NbSe$_2$ side, in addition to its normal $3\times3$ CDW (Fig.~\ref{fig:fig3}h--j, the constant-current d$I$/d$V$ maps). This additional modulation could arise from interference of incident and elastically scattered CDWs or from Friedel oscillations \cite{Zhang2022,Chen2023,Crommie1993} and it is absent on the VSe$_5$---NbSe$_6$ interfaces.

\section{Signatures of Kondo resonances in a side-coupled geometry at the interfaces.}
\begin{figure}[t!]
\centering
\includegraphics[width=1\textwidth]{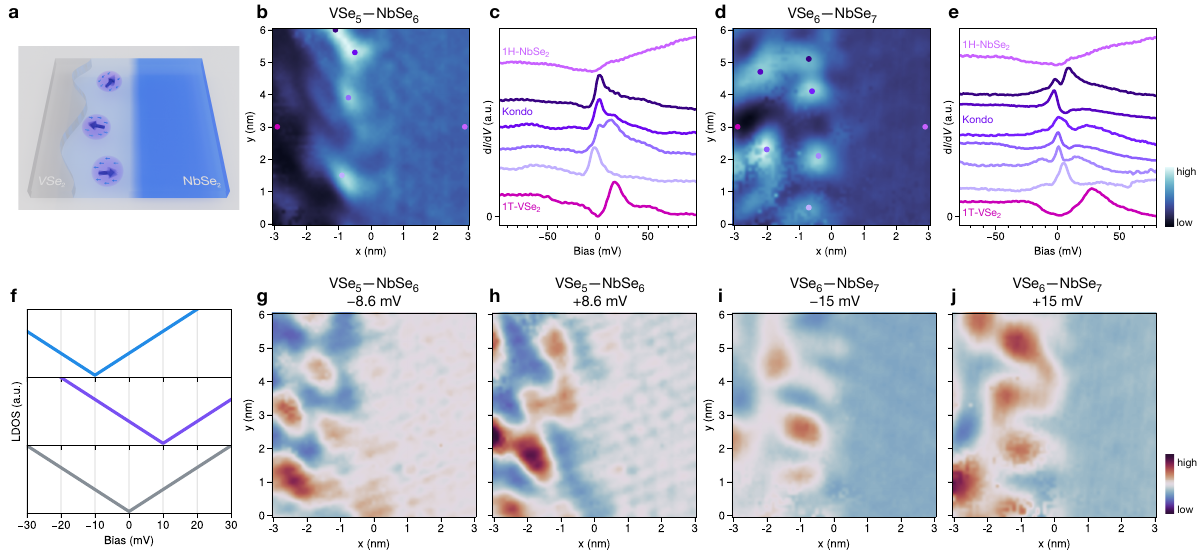}
\caption{\label{fig:fig4} Signatures of the Kondo effect in a side-coupled geometry and contrast inversion near the interfaces.
\textbf{a}, Schematic of Kondo singlets in a side-coupled geometry. Conduction electrons (light blue area) screen localized moments (purple arrows) forming Kondo singlets.
\textbf{b}, d$I$/d$V$ map at $-$0.8 mV of VSe$_5$---NbSe$_6$ interfaces, $V_\text{mod}$=2 mV. 
\textbf{c}, Point d$I$/d$V$ spectra from positions marked in \textbf{b} (shifted vertically for clarity).
\textbf{d}, d$I$/d$V$ map at 0 mV of VSe$_6$---NbSe$_7$ interfaces, $V_\text{mod}$=1 mV.
\textbf{e}, Point d$I$/d$V$ spectra from positions marked in \textbf{d} (shifted vertically for clarity).
\textbf{f}, Schematic of the V-shaped LDOS around Fermi level of different spots inside 1T-VSe$_2$: negatively charged (top), positively charged (middle) and neutral without charge doping (bottom). 
\textbf{g}, \textbf{h}, Contrast inversion of a d$I$/d$V$ map of VSe$_5$---NbSe$_6$ at ±8.6 mV, $V_\text{mod}$=2 mV. 
\textbf{i}, \textbf{j}, Contrast inversion of a d$I$/d$V$ map of VSe$_6$---NbSe$_7$ at ±15 mV, $V_\text{mod}$=1 mV.
}
\end{figure}

Lateral heterostructures provide an ideal platform to create 1D interfaces which combine properties of different materials and allow direct visualization of created novel phenomena in real space. Our heteroepitaxy protocol ensures atomically sharp interfaces largely eliminating the undesirable interference from imperfections (\textit{e.g.} defect states). Here, we provide one example of such novel effects in our artificial composite materials in a side-coupled geometry. In addition to the electronic effects discussed in the previous section, we observe both interfaces exhibiting strong localized zero bias anomalies with typical peak or dip-peak features (Fig.~\ref{fig:fig4}c, e and Supplementary Fig.~S8 and Fig.~S9). At the same time, our STS lacks typical inelastic tunneling features that could be associated with spin-flip or magnon excitations \cite{Ternes2015,Choi2019,Ganguli2023}. We attribute these zero bias anomalies to Kondo resonances arising from coupling between localized magnetic moments in 1T-VSe$_2$ and conduction electrons in 1H-NbSe$_2$ (Fig.~\ref{fig:fig4}a). These Kondo resonances are unlikely to arise from the coupling between 1T-VSe$_2$ and the substrate HOPG, since zero bias anomalies in pristine 1T-VSe$_2$ are absent in our results (Fig.~\ref{fig:fig4}c, e and Supplementary Fig.~S8 and Fig.~S9) and other reports on pure monolayer 1T-VSe$_2$/HOPG system \cite{Liu2018-2,Wong2019}. 
The emergence of free magnetic moments in 1T-VSe$_2$ could be related to the ground state of the material itself, or to an interfacial effect, with contributions from \textit{e.g.} inhomogenous strain and charge transfer. Similar Kondo resonances have been previously observed close to the edges of 1T-VSe$_2$ islands on bulk 2H-NbSe$_2$ \cite{Shawulienu2020}. 
Thus, it is most likely that 1H-NbSe$_2$ acts as the electron bath/charge reservoir which couples to the localized magnetic moments inside 1T-VSe$_2$ \cite{Esters2017,Wong2019,Shawulienu2020,Yin2022}. These lateral heterostructures realize Kondo resonances in a novel side-coupled geometry (Fig.~\ref{fig:fig4}a), which has not been reported before in an atomic-scale system, only in mesoscopic side-coupled quantum dot experiments \cite{Grobis2007,Sasaki2009,Hur2015}. 

The maximum intensity of the Kondo signal is found inside 1T-VSe$_2$ rather than exactly at the interface (Fig.~\ref{fig:fig4}b, d). They can extend in some cases even up to $\sim$8 nm away from the interface into the VSe$_2$ side (see Supplementary Fig.~S9). Besides the appearance of localized moments in VSe$_2$, the formation of these Kondo singlets depends on the possible spacial range of screening by the conduction electrons of NbSe$_2$. The positions of Kondo sites show real-space modulation, but they are not directly linked with the periodicity of either the VSe$_2$ lattice, or the CDW in 1T-VSe$_2$ or 1H-NbSe$_2$ (Fig.~\ref{fig:fig4}b,d) which could be related to the inhomogeneous charge distribution discussed below. This, taken together with the signature of Kondo resonances, suggests lack of magnetic order in 1T-VSe$_2$ around the interfaces. 

For 1T-VSe$_2$ near the interface, we also observe a contrast inversion in differential conductance (d$I$/d$V$) maps at small positive and negative bias, at energies inside the V-shaped local density of states (LDOS) (see Fig.~\ref{fig:fig4}f--j and Supplemenraty Fig.~S9). We attribute this phenomenon to spontaneous inhomogeneous electronic charge redistribution in 1T-VSe$_2$. Hence, at different positions the electronic doping shifts the 1T-VSe$_2$'s V-shaped LDOS towards positive/negative energy, as suggested in Fig.~\ref{fig:fig4}f. When obtaining d$I$/d$V$ maps \textit{e.g.} at positive bias +10 mV, the negatively charged spots get brighter contrast, while positively charged spots get darker contrast; and when taking at opposite bias, the contrast of d$I$/d$V$ maps reverses \cite{Spera2020,Pasztor2021}. In comparison, 1H-NbSe$_2$ is basically free from this contrast inversion. We speculate that this spontaneous charge redistribution may be a feature of 1T-VSe$_2$ itself, or a result of charge transfer to/from 1H-NbSe$_2$, which would depend on the CDW periodicity of both 1H-NbSe$_2$ and 1T-VSe$_2$, and thus could be quite inhomogenous.

\section{Conclusions}
In this work, we introduce a one-pot two-step heteroepitaxy method for constructing atomically sharp 1T-VSe$_2$---1H-NbSe$_2$ lateral heterostructures. We systematically study these defect-free, straight 1D interfaces to reveal their atomic-level geometric and electronic structures using STM and STS experiments corroborated by DFT calculations. We identify two structures, VSe$_5$---NbSe$_6$ and VSe$_6$---NbSe$_7$, and find additional electronic states and charge modulation localized at or near the 1D interfaces. We demonstrate that these types of lateral heterostructures can be used to realize novel electronic states, in our case Kondo resonances arising from the coupling of the magnetic moments of 1T-VSe$_2$ with the 1H-NbSe$_2$ conduction electrons. Our work presents a general method for constructing atomically perfect 1D interfaces in TMDC lateral heterostructures for further studies of correlated 1D systems.

\section{Methods}

\subsection{Experimental}
We use an enhanced MBE protocol (one-pot two-step molecular beam lateral epitaxy) to synthesize VSe$_{2}$, NbSe$_{2}$ and their lateral heterostructures (LH) in ultra-high vacuum (UHV) (base pressure of $\sim 8 \times 10^{-9}$ mbar.
Vanadium rod (99.8\%, Goodfellow Cambridge Ltd.) and niobium rod (99.9\%, MaTecK GmbH) were evaporated by electron-beam heating (EFM 3T, Focus GmbH). Se powder (99.99\%, Sigma-Aldrich) was evaporated in an effusion cell at $\sim$140 \textdegree C with a thermal cracker at $\sim$1200 \textdegree C (Thermal cracker cell, MBE-Komponenten GmbH). All samples were grown on highly oriented pyrolytic graphite (HOPG) (ZYB grade, TipsNano Co.), which were previously degassed above $\sim$600 \textdegree C. The VSe$_{2}$---NbSe$_{2}$ lateral heterostructure was synthesized in two steps: first growing NbSe$_{2}$ at $\sim$550 \textdegree C with a growth rate of 29 min per monolayer, with 30 min post annealing at $\sim$400 \textdegree C; then second growing VSe$_{2}$ at a substrate temperature of $\sim$375 \textdegree C with a growth rate of 14 min per monolayer, with 5 min post annealing at $\sim$ 375 \textdegree C. Later, to protect materials during transferring to STM, samples were capped in Se vapor with an amorphous Se layer ($>10$ nm). The STM experiments were carried out in another UHV setup following removal of the selenium capping by gentle thermal annealing of the samples ($<300$ \textdegree C). All the STM images and spectra were acquired at $\sim$ 4.2 K using a Createc LT-STM (CreaTec GmbH), or Unisoku USM-1300 (Unisoku Co., Ltd) at 2 K (Fig.~1d). For STM topography images feedback set-point bias voltage $V_\text{s}$, tunnelling current ($I_\text{t}$) are given in the figure captions. Topography images are rendered with Gwyddion \cite{Gwyddion2012}. Scanning tunnelling spectra (STS) or differential conductance (d\textit{I}/d\textit{V}) spectroscopy are measured with lock-in technique at the frequency of 746 Hz, and the peak-to-peak modulation voltage ($V_\text{mod}$) are specified in the figure caption. 

\subsection{Density functional theory calculations}
DFT calculations were performed with the \textit{QUANTUM ESPRESSO} distribution \cite{Giannozzi2009}. Interaction between electrons and ions were described with the PAW pseudopotentials \cite{PhysRevB.83.195131,PhysRevLett.92.246401}, while the electronic wave functions were expanded considering a plane-wave basis set with kinetic energy cutoffs of 90 Ry. For the lateral heterostructures, the integration over the Brillouin zone (BZ) were performed using a uniform grid of 1x8x1 k-point. For the results shown in the main paper, we have adopted the standard Perdew-Burke-Ernzerhof (PBE) functional augmented with +U correction of 2 eV on V-3d orbitals \cite{PhysRevLett.77.3865,PhysRevB.71.035105}. Results with other values of +U are present in the SI. STM simulations and LDOS maps were obtained with the \textit{critic2} code \cite{critc2_1,critc2_2}.

\section{Contributions}
X.H. and H.G.H. contributed equally to this work. X.H., S.K. and P.L. conceived the idea and designed the experiment. X.H. and H.G.-H. assembled the MBE and synthesized materials, and performed STM measurements. O.J.S. and X.H. performed DFT calculations. X.H. analyzed STM data. X.H., H.G.-H., J.S. and P.L. composed the manuscript and all authors discussed the results and commented on the manuscript.

\section{Acknowledgements}
This research made use of the Aalto Nanomicroscopy Center (Aalto NMC) facilities and was supported by the European Research Council (ERC-2017-AdG no.~788185 ``Artificial Designer Materials'') and Academy of Finland (Academy professor funding nos.~318995 and 320555, Academy research fellow nos.~338478 and 346654). Computing resources from the Aalto Science-IT project and CSC, Helsinki are gratefully acknowledged. X.H. thanks Mr. HUANG Ruojun and Mrs. XIONG Dongyan. H.G-H. acknowledges financial support from the Spanish State Research Agency under grant Ramón y Cajal fellowship RYC2021-031050-I.

\bibliography{refs}

\end{document}

\title{Supplementary Information: \\Atomically sharp 1D interfaces in 2D lateral heterostructures \\of VSe$_2$---NbSe$_2$ monolayers}
\author{Xin Huang}
\thanks{These two authors contributed equally}

\affiliation{Department of Applied Physics, Aalto University, FI-00076 Aalto, Finland}

\author{H\'ector Gonz\'alez-Herrero}
\thanks{These two authors contributed equally}

\affiliation{Department of Applied Physics, Aalto University, FI-00076 Aalto, Finland}
\affiliation{Departamento F\'isica de la Materia Condensada, Universidad Aut\'onoma de Madrid, Madrid E-28049, Spain}

\author{Orlando J. Silveira}
\affiliation{Department of Applied Physics, Aalto University, FI-00076 Aalto, Finland}

\author{Shawulienu Kezilebieke}
\affiliation{Department of Physics, Department of Chemistry and Nanoscience Center, 
University of Jyväskyl\"a, FI-40014 University of Jyväskyl\"a, Finland}

\author{Peter Liljeroth}
\affiliation{Department of Applied Physics, Aalto University, FI-00076 Aalto, Finland}

\author{Jani Sainio}
\affiliation{Department of Applied Physics, Aalto University, FI-00076 Aalto, Finland}

\maketitle

%\Xin{dIdV map at various bias, from scan directly, not from grid}

\section{Experimental section}
\linenumbers
\begin{figure}[h!]
\centering
\includegraphics[width=1\textwidth]{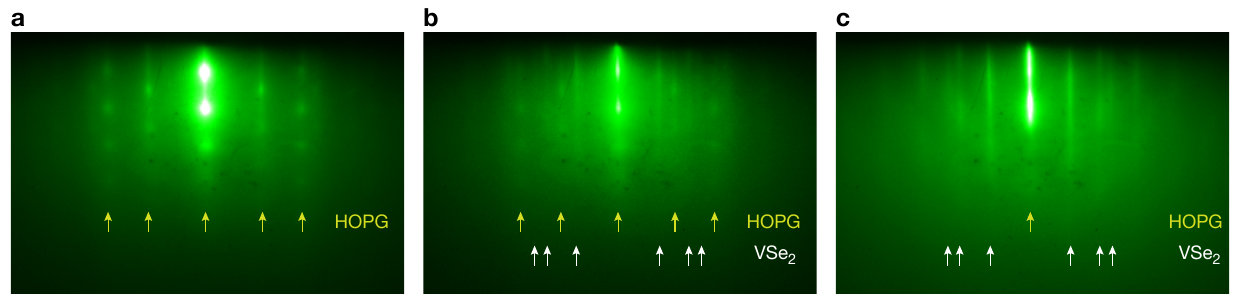}
\caption{\label{}
Reflection high-energy electron diffraction (RHEED) pattern during the growth of monolayer VSe$_2$.
\textbf{a}, Before the growth, RHEED pattern of the HOPG substrate.
\textbf{b}, During the growth. VSe$_2$ stripes start appearing.
\textbf{c}, End of growth. VSe$_2$ now has a high coverage, HOPG's pattern has diminished.
}
\end{figure}

\begin{figure}[h!]
\centering
\includegraphics[width=0.7\textwidth]{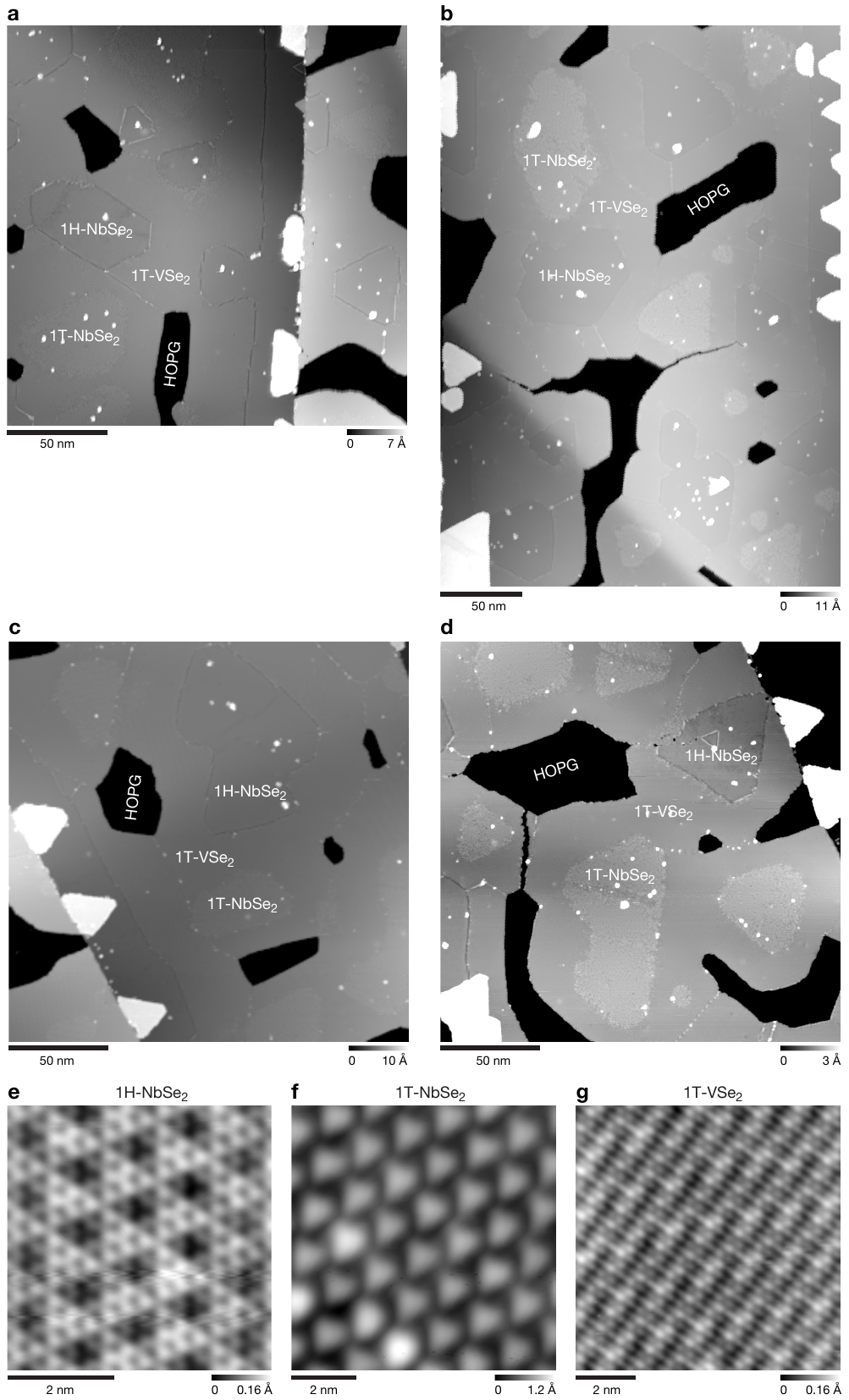}
\caption{\label{}
Lateral heteroepitaxy protocol: VSe$_2$ preferentially grows laterally from the edges of NbSe$_2$ islands, rather than exhibting vertical growth. During the NbSe$_2$ growth step, both phases 1H- and 1T-NbSe$_2$ can be found and consequently, we form lateral heterostructures of the two NbSe$_2$ phases with 1T-VSe$_2$.
\textbf{a}--\textbf{d}, Large-scale STM topography images of several different regions of the sample. Typical charge density wave of \textbf{e}, 1H-NbSe$_2$; \textbf{f}, 1T-NbSe$_2$; \textbf{g}, 1T-VSe$_2$, from STM topography image.
Scan parameters:
\textbf{a}, $V_\text{s}$=+1.36 V, $I_\text{t}$=10 pA. 
\textbf{b}, $V_\text{s}$=$-$1.5 V, $I_\text{t}$=10 pA.
\textbf{c}, $V_\text{s}$=+1.541 V, $I_\text{t}$=6.5 pA.
\textbf{d}, $V_\text{s}$=+1.5 V, $I_\text{t}$=4.1 pA.
\textbf{e}, $V_\text{s}$=$-$0.36 V, $I_\text{t}$=100 pA.
\textbf{f}, $V_\text{s}$=$-$0.649 V, $I_\text{t}$=400 pA.
\textbf{g}, $V_\text{s}$=$-$1 V, $I_\text{t}$=100 pA.
}
\end{figure}

\begin{figure}[h!]
\centering
\includegraphics[width=1\textwidth]{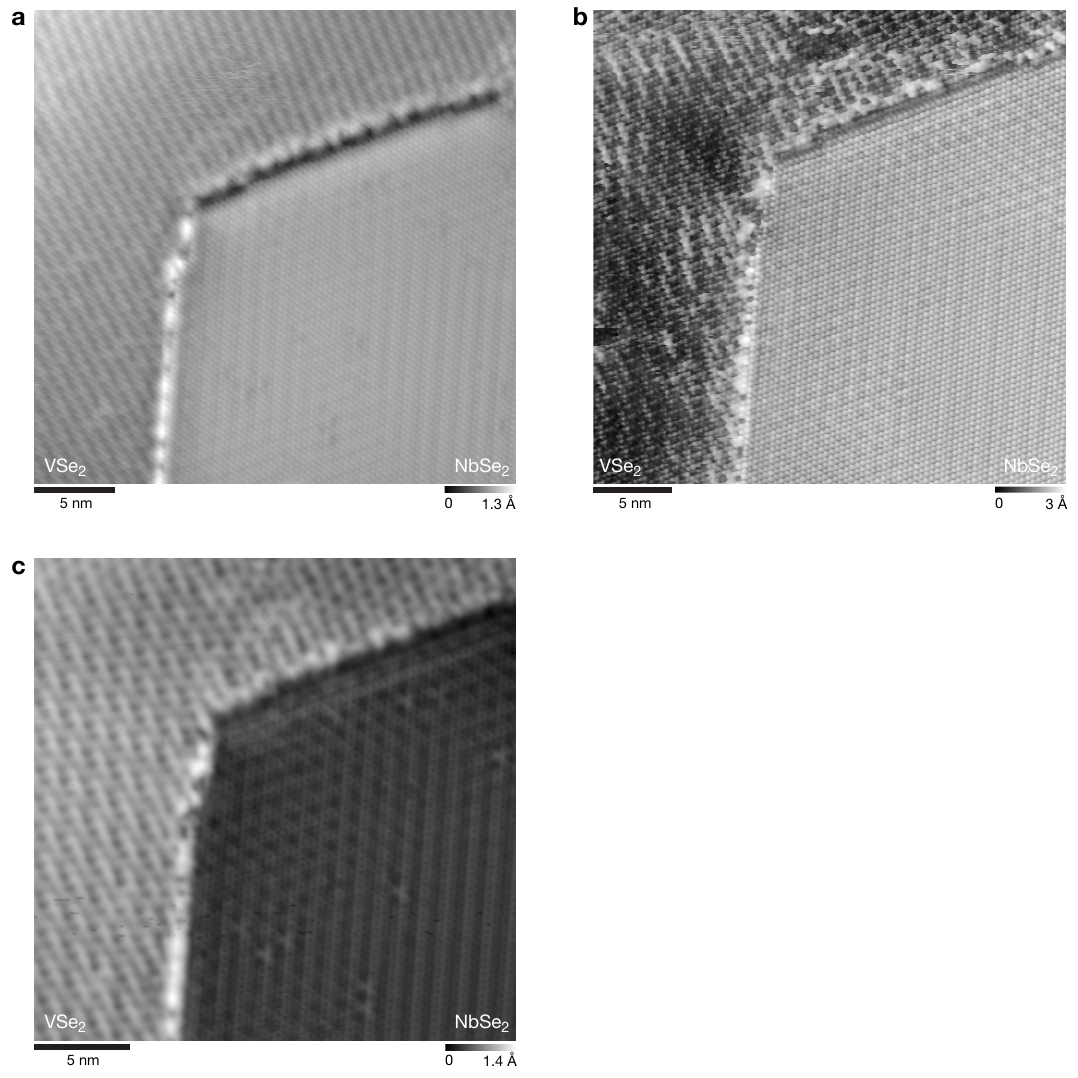}
\caption{\label{} Two adjacent lateral heterostructures at corner of a NbSe$_2$ island (on A-side). The interfaces shows zigzag morphology, with one of them having bright and and the other dark contrast. The intrinsic CDWs of both materials extend right up to the interface; neither the commensurate CDW of 1H-NbSe$_2$ nor the incommensurate CDW of 1T-VSe$_2$ extends into the other material. Both interfaces can extend up to $\sim$ 20 nm.
\textbf{a}, Topography image of Fig.~2\textbf{b}, \textbf{i}, \textbf{j} and Fig.~3\textbf{h}--\textbf{j} ($V_\text{s}$=$-$1 V, $I_\text{t}$= 100 pA).
\textbf{b}, Extended data of topography image with atomic resolution ($V_\text{s}$=$-$10 mV, $I_\text{t}$= 790 pA).
\textbf{c}, Extended data of topography image with charge density wave ($V_\text{s}$=$-$201 mV, $I_\text{t}$= 50 pA).
}
\end{figure}

\begin{figure}[h!]
\centering
\includegraphics[width=0.8\textwidth]{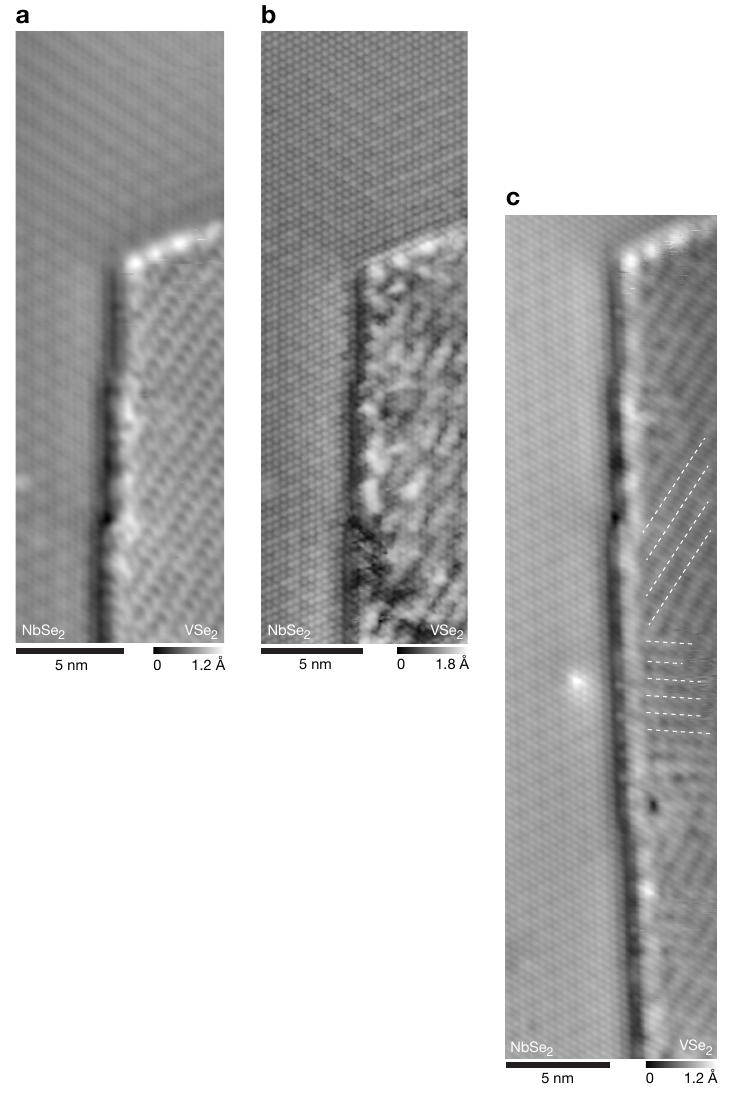}
\caption{\label{} Additional STM topography images of two adjacent lateral heterostructures across a corner of a NbSe$_2$ island, on A-side. Panel \textbf{a}--\textbf{c} are aligned with the same position of heterostructures.
\textbf{a}, STM topography image ($V_\text{s}$=$-$1 V, $I_\text{t}$= 100 pA).
\textbf{b}, STM topography image with atomic resolution ($V_\text{s}$=$-$10 mV, $I_\text{t}$= 1 nA). 
\textbf{c}, STM topography image. The interface can extend up to $\sim$ 40 nm. In the middle part of the dark interface, although the CDW of 1T-VSe$_2$ changes its direction (indicated with dashed lines), the dark contrast of the interface does not change. ($V_\text{s}$=$-$1 V, $I_\text{t}$= 100 pA).
}
\end{figure}

\begin{figure}[h!]
\centering
\includegraphics[width=0.8\textwidth]{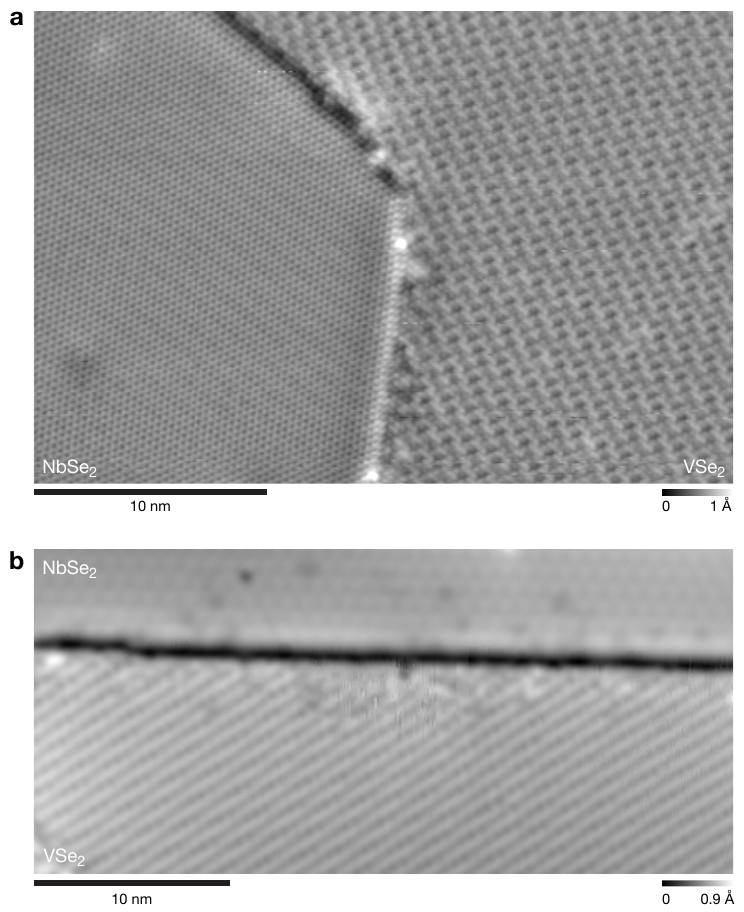}
\caption{\label{} 
\textbf{a}, Additional STM topography image of lateral heterostructures across a corner of a NbSe$_2$ island, on A-side. ($V_\text{s}$=$-$1.503 V, $I_\text{t}$=31 pA).
\textbf{b}, Additional STM topography image of a long interface $\sim$ 36 nm ($V_\text{s}$=+1.36 V, $I_\text{t}$=90 pA). 
}
\end{figure}

\begin{figure}[h!]
\centering
\includegraphics[width=1\textwidth]{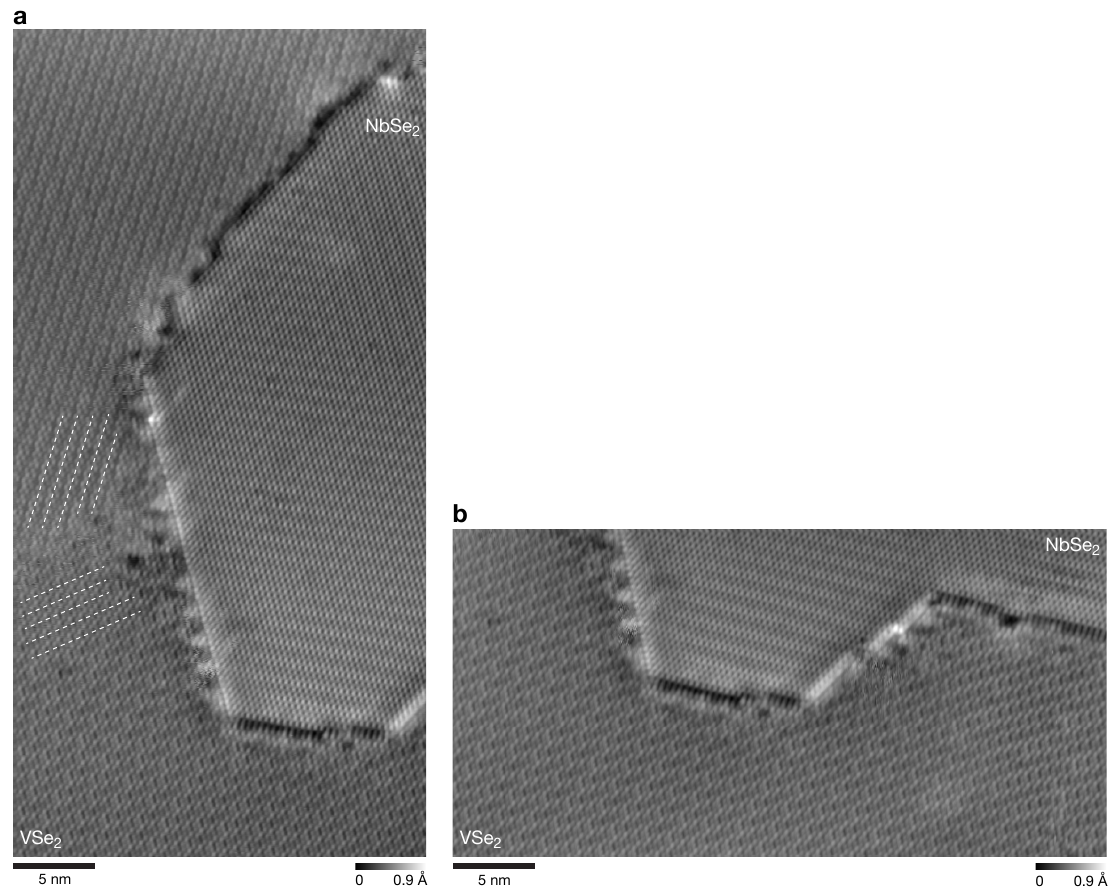}
\caption{\label{} Adjacent lateral heterostructures across corners of a NbSe$_2$ island, on B-side. The interfaces shows parallel morphology, and bright and dark contrast appears alternatingly at adjacent edges.
\textbf{a}, Topography image of Fig.~2\textbf{d}, \textbf{l}. Along the bright interface, in the middle, the CDW of 1T-VSe$_2$ changes direction (indicated with dashed lines) but the bright contrast doesn't change. ($V_\text{s}$=$-$1 V, $I_\text{t}$= 100 pA).
\textbf{b}, Extended data of topography image ($V_\text{s}$=$-$1 V, $I_\text{t}$= 100 pA).
}
\end{figure}

\begin{figure}[h!]
\centering
\includegraphics[width=1\textwidth]{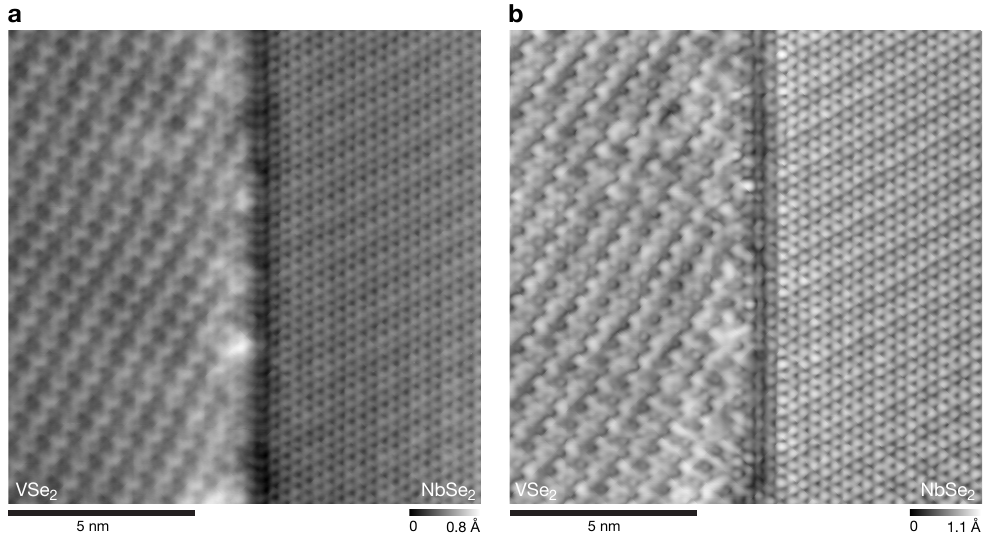}
\caption{\label{}
VSe$_5$---NbSe$_6$ lateral heterostructure, on B-side.
\textbf{a}, Topography image of the interface shown in Fig.~2\textbf{k}. ($V_\text{s}$=$-$0.99 V, $I_\text{t}$= 50 pA).
\textbf{b}, Extended data of topography image ($V_\text{s}$=$-$50 mV, $I_\text{t}$= 200 pA).
}
\end{figure}

\begin{figure}[h!]
\centering
\includegraphics[width=0.85\textwidth]{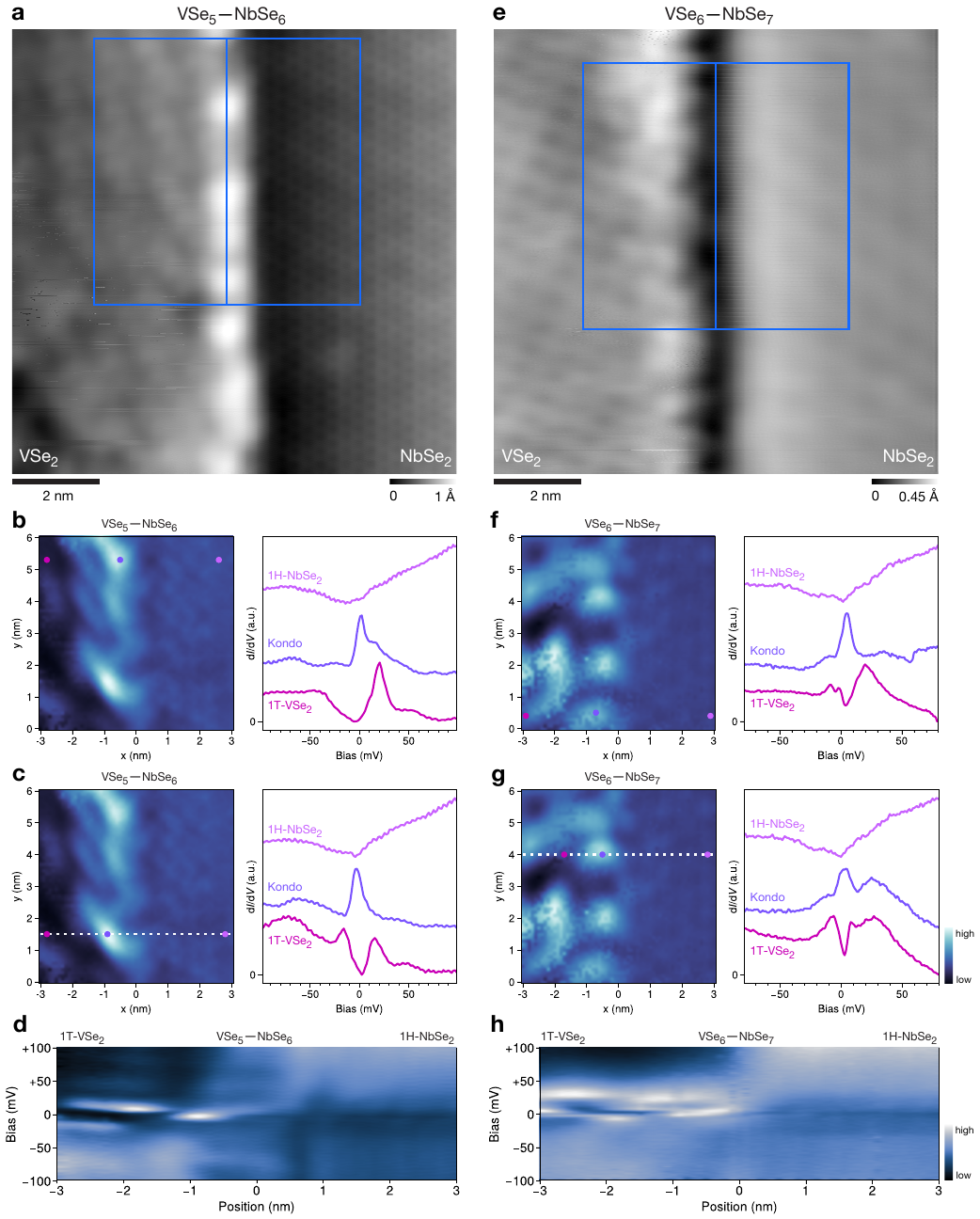}
\caption{\label{}
Signatures of Kondo resonances in a side-coupled geometry.
\textbf{a}, \textbf{e}, STM topography of VSe$_5$---NbSe$_6$ (Fig.~4 and Fig.~3\textbf{g}) and VSe$_6$---NbSe$_7$ interfaces (Fig.~4). Blue square frames indicate the d$I$/d$V$ map area; the middle line indicates the 0 position of x-axis in those d$I$/d$V$ maps. We do not observe defects or alloying in the mapping area.
\textbf{b}, \textbf{c}, (left) d$I$/d$V$ map at $-$0.8 mV of VSe$_5$---NbSe$_6$ interfaces ($V_\text{mod}$=2 mV), and (right) corresponding point d$I$/d$V$ spectra (positions marked in left panel; spectra are shifted vertically for clarity).
\textbf{d}, d$I$/d$V$ spectra along the line indicated in \textbf{c}.
\textbf{f}, \textbf{g}, (left) d$I$/d$V$ map at 0 mV of VSe$_6$---NbSe$_7$ interfaces ($V_\text{mod}$=2 mV), and (right) corresponding point d$I$/d$V$ spectra (positions marked in left panel; spectra are shifted vertically for clarity).
\textbf{h}, d$I$/d$V$ spectra along the line indicated in \textbf{c}.
Scan parameters:
\textbf{a}, $V_\text{s}$=$-$0.5 V, $I_\text{t}$=210 pA. 
\textbf{e}, $V_\text{s}$=$-$1.5 V, $I_\text{t}$=380 pA. 
}
\end{figure}

\begin{figure}[h!]
\centering
\includegraphics[width=0.63\textwidth]{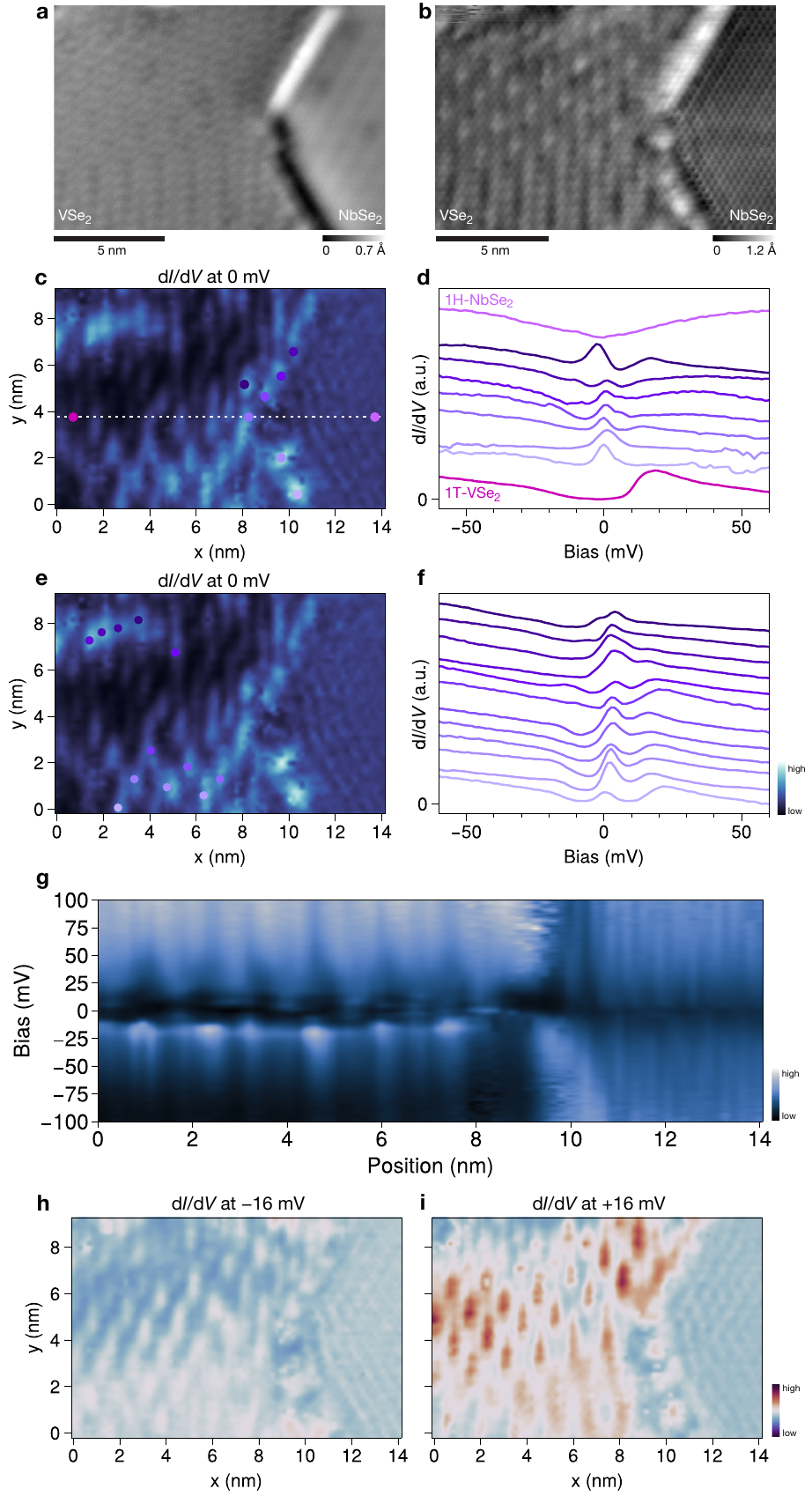}
\caption{\label{}
Two lateral heterostructures and Kondo resonances in a side-coupled geometry.
\textbf{a}, STM topography. $V_\text{s}$=+1.5 V, $I_\text{t}$=80 pA.
\textbf{b}, STM topography with atomic resolution. $V_\text{s}$=$-$0.1 V, $I_\text{t}$= 79 pA.
\textbf{c}, d$I$/d$V$ map at 0 mV ($V_\text{mod}$=2 mV).
\textbf{d}, d$I$/d$V$ spectra of corresponding points in \textbf{c}, with points in 1T-VSe$_2$ and 1H-NbSe$_2$ (spectra are shifted vertically for clarity).
\textbf{e}, d$I$/d$V$ map at 0 mV ($V_\text{mod}$=2 mV). Signatures of Kondo resonances can be found in VSe$_2$ up to $\sim$8 nm away from the interfaces.
\textbf{f}, d$I$/d$V$ spectra of corresponding points in \textbf{e} (spectra are shifted vertically for clarity).
\textbf{g}, d$I$/d$V$ spectra along the line indicated in \textbf{c}.
\textbf{h} and \textbf{i}, Contrast inversion of a dI/dV map of these heterostructures at ±16 mV.
}
\end{figure}

\begin{figure}[h!]
\centering
\includegraphics[width=1\textwidth]{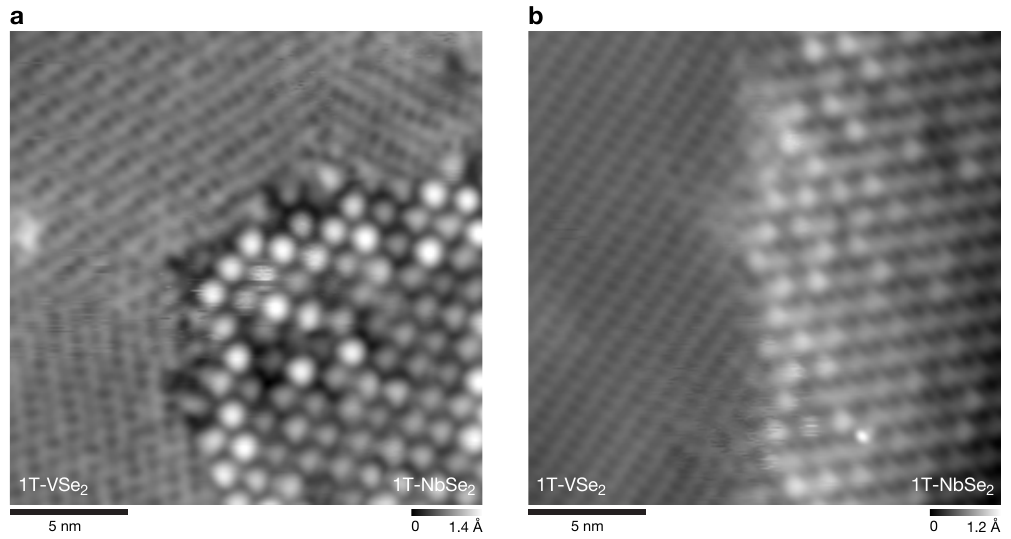}
\caption{\label{}
Additional STM topography of lateral heterostructures of 1T-VSe$_2$ and 1T-NbSe$_2$ islands. Also in this case neither CDW extends into the other material. Scan parameters: \textbf{a}, $V_\text{s}$=$-$0.247 V, $I_\text{t}$=62 pA. 
\textbf{b}, $V_\text{s}$=$-$1.5 V, $I_\text{t}$=150 pA. 
}
\end{figure}

\section{DFT calculations}

\subsection{Electronic properties of 2D 1H-NbSe$_2$ and 1T-VSe$_2$}

\begin{figure}[h!]
\centering
\includegraphics[width=.75\textwidth]{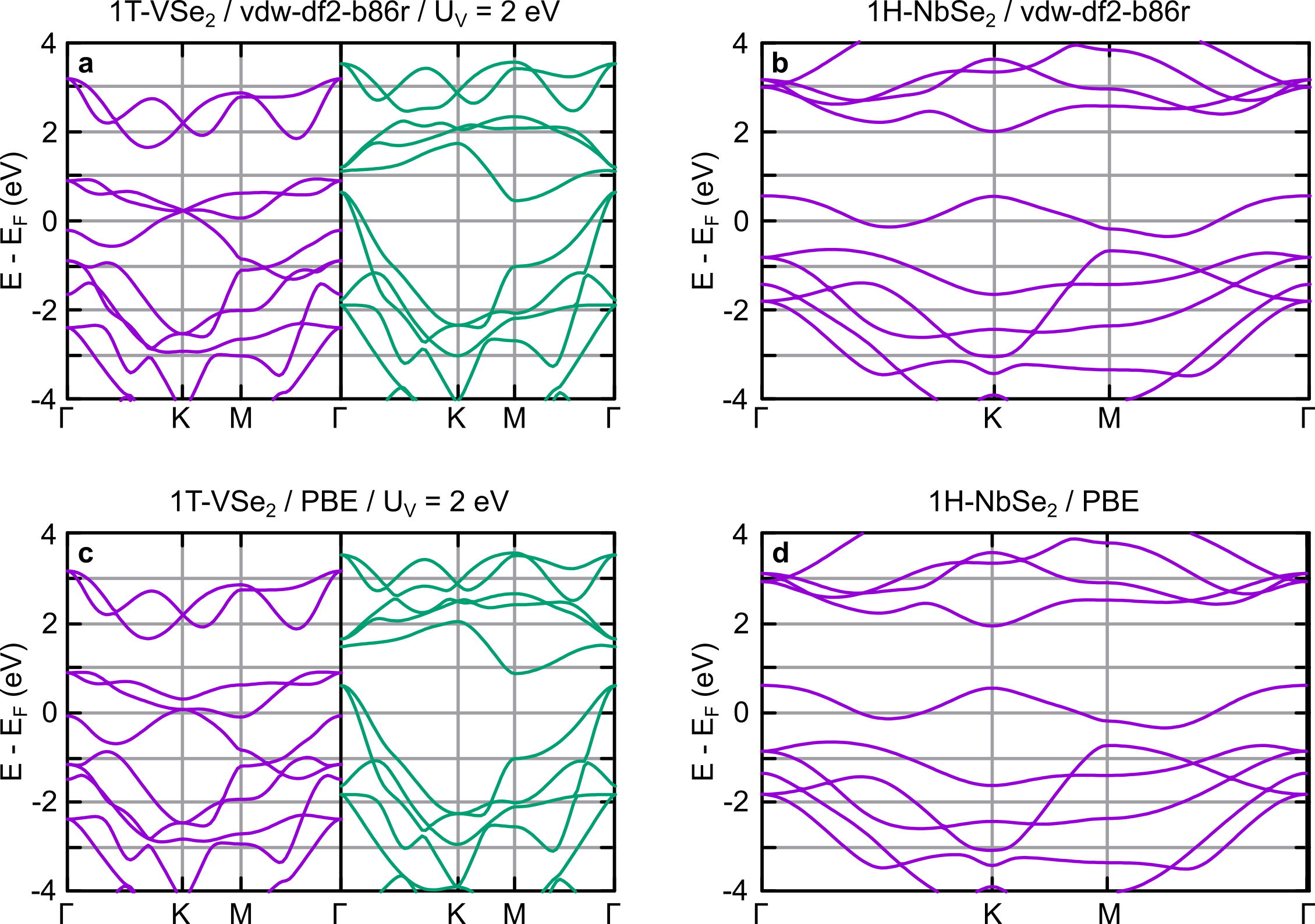}
\caption{\label{dft:bulkbands}DFT-calculated band structures of 2D 1H-NbSe$_2$ and 2D 1T-VSe$_2$ obtained with different functionals}
\end{figure}

\begin{figure}[h!]
\centering
\includegraphics[width=.75\textwidth]{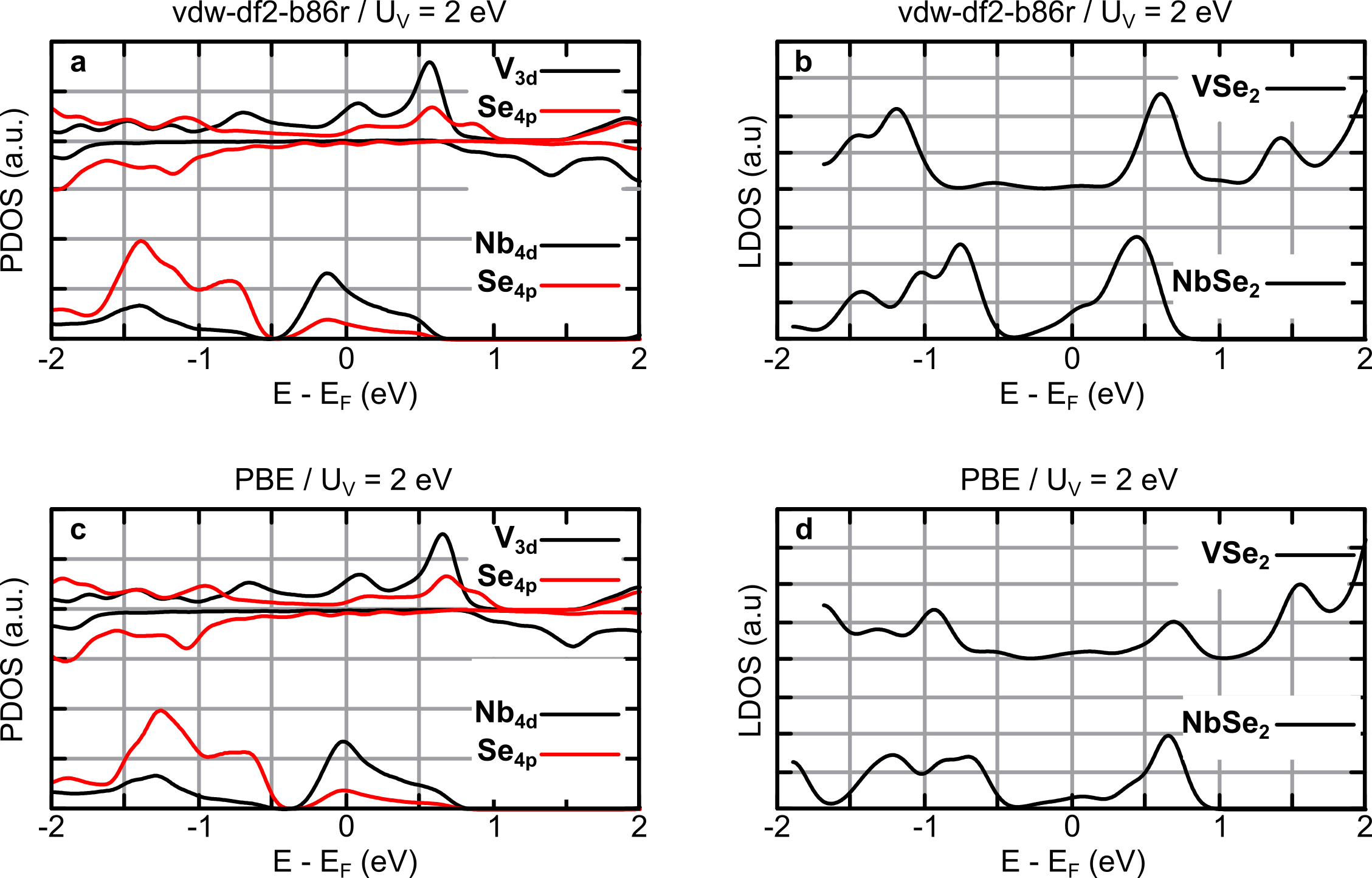}
\caption{\label{dft:bulkpdos}DFT-calculated PDOS and simulated STS calculated by integrating the local density of states (LDOS) at a constant height of 3 Å over the topmost atom of each structure.}
\end{figure}

We tested the effect of the vdw-df2-b86r functional \cite{PhysRevB.89.121103} to the electronic properties of 2D 1H-NbSe$_2$ and 2D 1T-VSe$_2$. It is known that the vdw-df2-b86r functional with moderate U (e.g. 2 eV) is more appropriate to treat 1T-VSe$_2$ \cite{Shawulienu2020} as opposed to standard functionals such as PBE. Indeed, our DFT results in Figure \ref{dft:bulkbands} reveal that the vdw-df2-b86r gives minor quantitative changes in the 1T-VSe$_2$ band structure, specially at energies close to the Fermi level, while the band structure of 1H-NbSe$_2$ seems to be unaffected. Figure \ref{dft:bulkpdos} shows the projected density of states (PDOS) and the simulated scanning tunneling spectroscopy (STS), both calculated with the vdw-df2-b86r and PBE functionals for comparison. Apart from very small shifts, the PDOS and simulated STS calculated with different functionals present overall the same phenomenology, specially the simulated STS, which is the most important aspect needed in this work. Thereby, we used the PBE functional with U = 2 eV throughout the calculations with the lateral interface. 

\subsection{Structural and electronic properties of the interfaces}

\begin{figure}[h!]
\centering
\includegraphics[width=.9\textwidth]{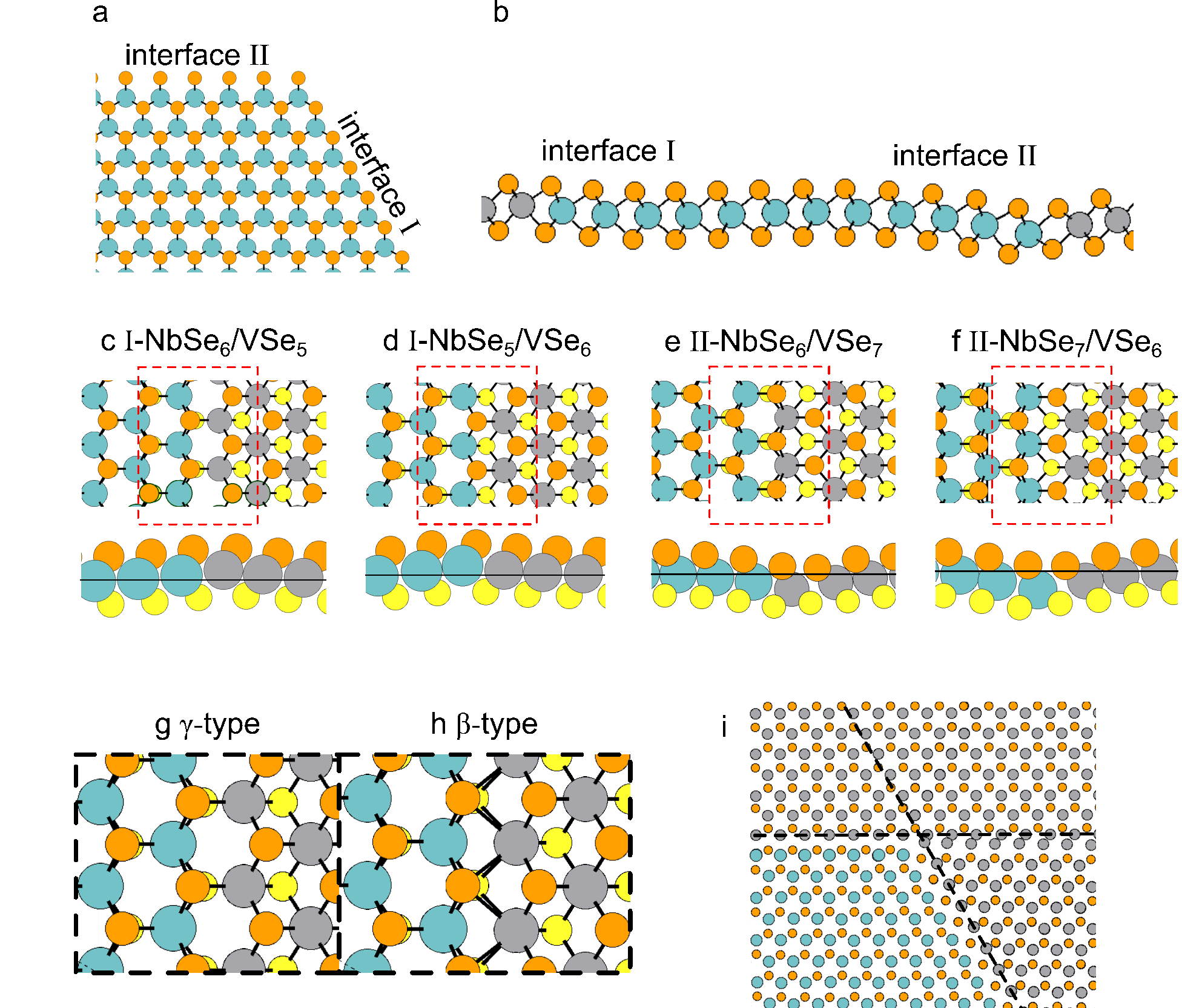}
\caption{\label{dft:energies}a: top view of an NbSe$_2$ island showing two possible edge configurations. b: Side view of an example of lateral heterostructures. c-f: Proposed interfaces considered based on the coordination number of the Nb and V atoms at the lateral interface (A-side). g and h: Examples of the $\gamma$ and $\beta$ types of interfaces. i: Example of a tentative overlapping of a heterostructure with coexisting two $\beta$-type interface, showing that it cannot grow in registry (only Nb, V and top Se atoms are shown).}
\end{figure}

\begin{figure}[h!]
\centering
\includegraphics[width=.9\textwidth]{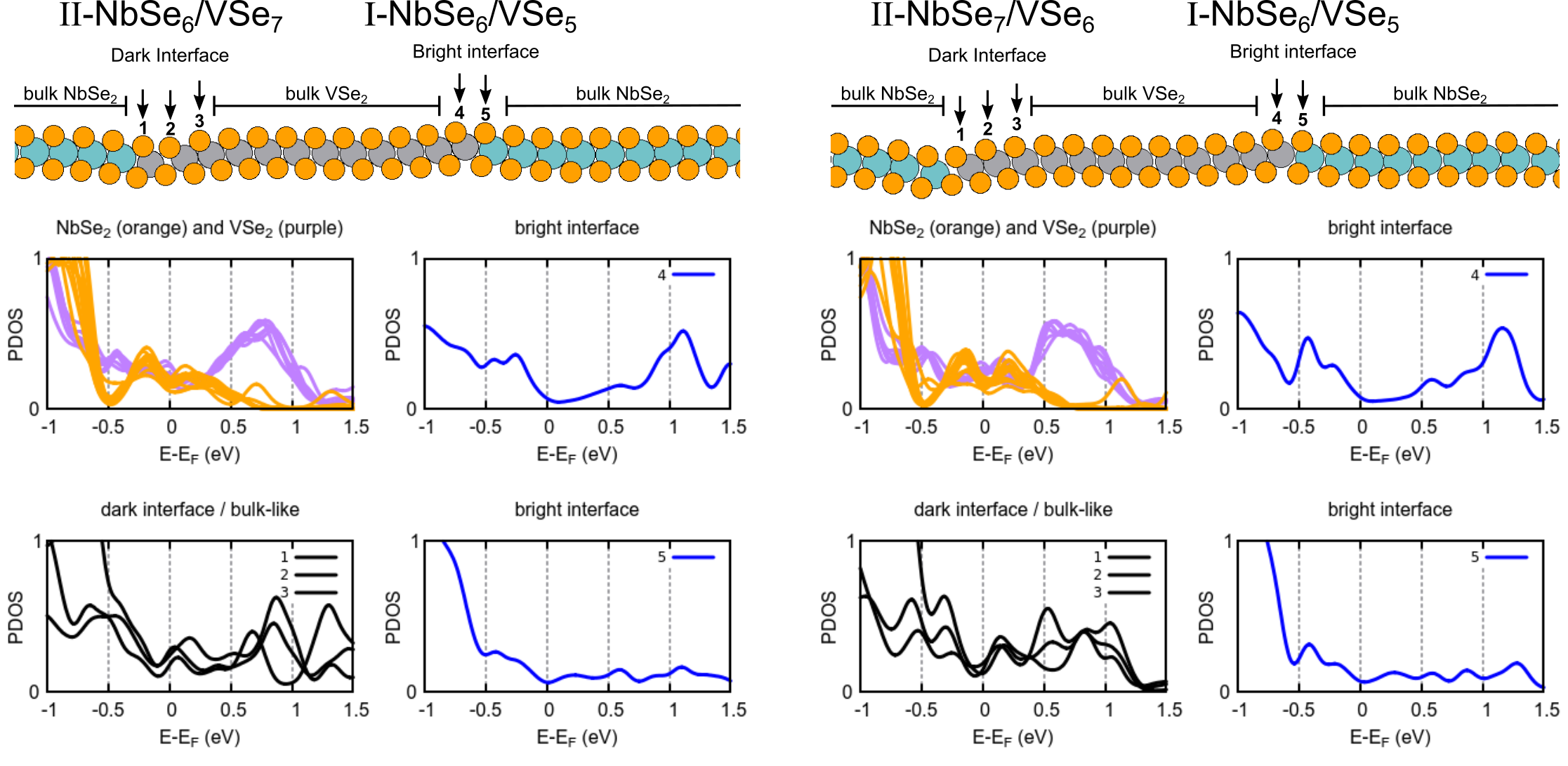}
\caption{\label{dft:singleatom}DFT-calculated PDOS on the $p_z$
orbitals of interface region Se atoms (A-side).}
\end{figure}

Considering the PBE(U=2 eV) DFT setup defined in the previous section, we performed full relaxation and total energy calculations of different interface structures based on the edge scenarios I and II for the NbSe$_2$ island shown in Figure \ref{dft:energies}a. Given the nature of the edges in the NbSe$_2$ islands, the edges I and II can coexist in the same unit cell, for example, in the lateral heterostructure shown in Figure \ref{dft:energies}b, and four NbSe$_2$/VSe$_2$ interface candidates that can grow in registry were created based on the coordination number of either Nb or V atoms at the interface: I-NbSe$_6$/VSe$_5$, I-NbSe$_5$/VSe$_6$, II-NbSe$_6$/VSe$_7$ and II-NbSe$_7$/VSe$_6$ (Figures \ref{dft:energies}c-f, respectively). Notice that all four interfaces present the zig-zag pattern (top orange Se atoms) and parallel pattern (bottom yellow Se atoms) simultaneously. In order to investigate which interfaces in the lateral heterostructure are more energetically favorable among the structures proposed here, we compared the total energy of structures that differ by only one of the interfaces, thus the total number of atoms will be the same. For example, by keeping the same type of interface I, we were able to calculate and compare the total energies of the two types of interface II shown in Figure \ref{dft:energies}e and f, and the same procedure goes by keeping interface II as well. Our calculations show that among the type I interfaces, the I-NbSe$_6$/VSe$_5$ is more stable than the I-NbSe$_5$/VSe$_6$ by 0.2 eV, while among type II the II-NbSe$_7$/VSe$_6$ is more stable than II-NbSe$_6$/VSe$_7$ by 0.4 eV. Even though the total energies are affected by a twist of the unit cell that occurs in order to keep two interfaces, the energy spent in this deformation is two orders of magnitude smaller than the energy differences obtained here, thus not creating any artificial stability. We also stress that all interfaces considered here are of the $\gamma-$type, whereas interfaces of the $\beta-$type do not grow in registry and would induce strong strain and reorganization of the interfaces, which is not compatible to the experimental results (see Figures \ref{dft:energies}g, h and i). Mixing of $\gamma$ and $\beta$-types also cannot grow in registry. A simple way to distinguish the $\beta$- and $\gamma$-type interfaces is if transition metal atoms are aligned or not: for $\beta$-type, metal atoms are aligned in parallel, while for $\gamma$-type metal atoms are in a zigzag configuration at the interfaces. \cite{Lin2014}

The DFT-calculated energies indicate that the islands where the interfaces I-NbSe$_6$/VSe$_5$ and II-NbSe$_7$/VSe$_6$ coexist are preferable. However, even though the I-NbSe$_6$/VSe$_5$ is already expected to occur due to the stability of the edge I Nb having coordination six, it is very difficult to rule out the occurrence of the II-NbSe$_6$/VSe$_7$ interface just based on our STM and LDOS results. We then proceeded to investigate two lateral heterostructures shown in Figure \ref{dft:singleatom}, which differ only by the interface II. Additionally, the atomic contributions to the different STM signals and LDOS in the bright and dark interfaces are elucidated in more details in Figure \ref{dft:singleatom}, showing the DFT-calculated PDOS on the $p_z$ orbitals of each top Se atom (A-side structure) of both lateral heterostructures. These are the atoms that contribute the most to the tunneling current, and will directly influence the contrast in the STM and LDOS either due to an electronic effect, for example having lower or higher density of states, or due to the undulation of the relaxed structure. In both lateral heterostructures, the Se atom belonging to VSe$_2$ immediately at the bright interface has a significantly shifted PDOS compared to its surrounding Se atoms, which is then responsible to the strong bright features observed at $-$0.4 eV and 1.2 eV in the LDOS in the main text Figure 3f. Aside from a shift in energy, both features are consistent with the ones observed in the experiment at the bright interface. For the dark interface of the lateral heterostructure at the left of Figure \ref{dft:singleatom}, we observed a small attenuation of the PDOS of the Se at the interface belonging to the VSe$_2$, which are much lower in height than the other atoms in the bulk area. We then attribute in this case the darker contrast to both electronic and geometric effects. On the other hand, no attenuation is observed in the heterostructure at the right side of Figure \ref{dft:singleatom}, meaning that only the undulation is responsible to the dark contrast observed.

\begin{figure}[h!]
\centering
\includegraphics[width=.8\textwidth]{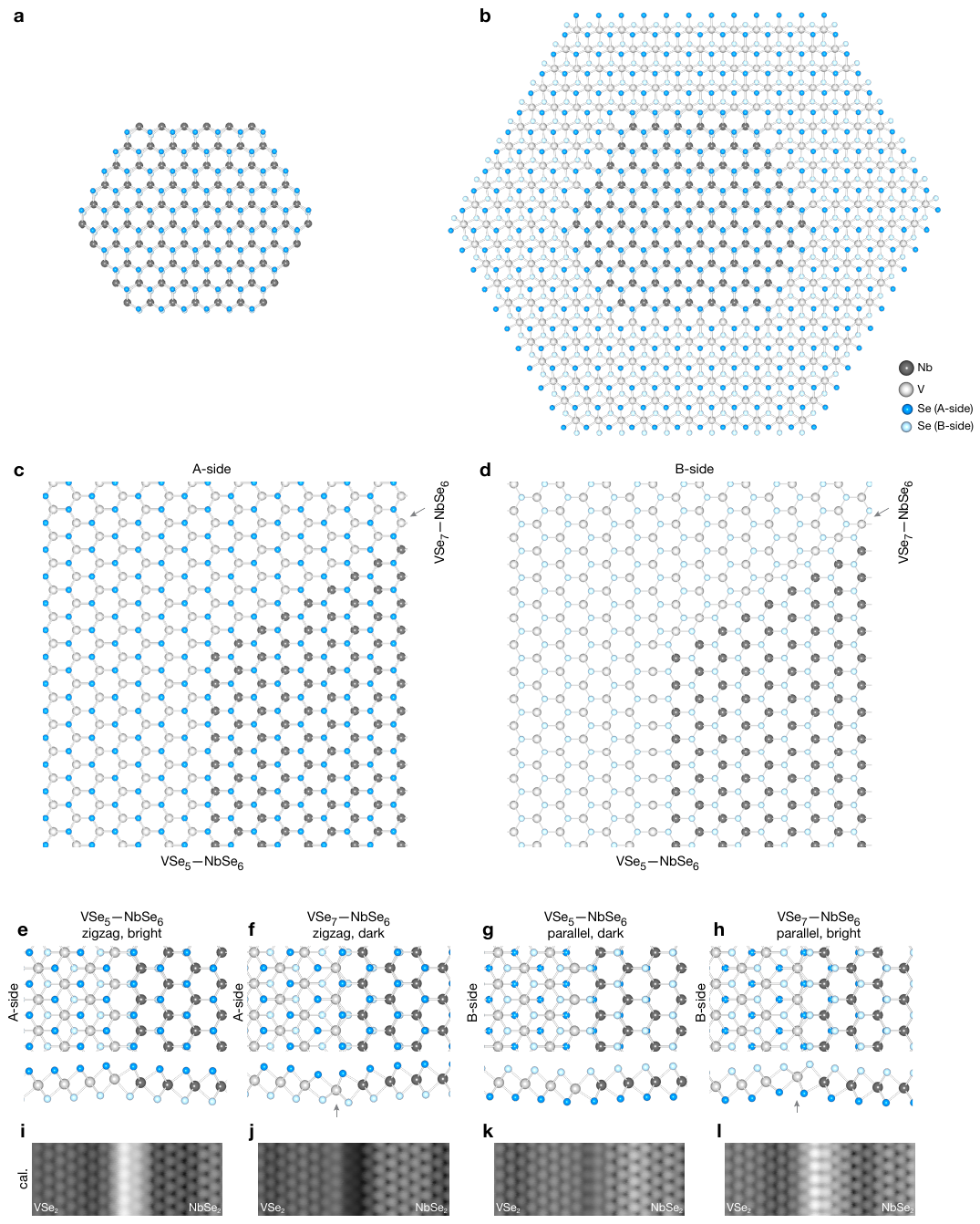}
\caption{\label{dft:alternative}Alternative lateral heterostructure geometry with VSe$_5$---NbSe$_6$ and VSe$_7$---NbSe$_6$ interfaces. Compared to results reported in main text, the structure of the VSe$_5$---NbSe$_6$ interface remains unaltered. The main structural difference between VSe$_7$---NbSe$_6$ and VSe$_6$---NbSe$_7$ is that a line of transition metal atoms at interface is replaced from V atoms to Nb atoms (indicated by the grey arrows in panel \textbf{c}, \textbf{d}, \textbf{f}, \textbf{h}).
\textbf{a}, \textbf{b}, Schematics of the formation of  a lateral heterostructure in this geometry: VSe$_2$ and NbSe$_2$ can also grow in-registry with VSe$_5$---NbSe$_6$ and VSe$_7$---NbSe$_6$ interfaces.
\textbf{c}, \textbf{d}, Zoom-in schematic of two interfaces with views from A-side (with A-side Se atoms only) and B-side (with B-side Se atoms only), respectively. Grey arrows indicate the difference between VSe$_7$---NbSe$_6$ and VSe$_6$---NbSe$_7$ (schematic structures without relaxation).
\textbf{e}--\textbf{h}, Top view and side view of the these lateral heterostructures of VSe$_5$---NbSe$_6$ and VSe$_7$---NbSe$_6$, 
and \textbf{i}--\textbf{l}, their corresponding calculated STM images ($V_\text{s}$=$-$0.5 V) (calculated structures with relaxation). Grey arrows indicate the difference between VSe$_7$---NbSe$_6$ and VSe$_6$---NbSe$_7$.
}
\end{figure}

Figure \ref{dft:alternative} shows the alternative lateral heterostructure where only interface II is different from the one presented in the main text. The simulated STM images for the two types of interface II structures (VSe$_7$---NbSe$_6$ / VSe$_6$---NbSe$_7$) are quite similar but as mentioned before VSe$_7$---NbSe$_6$ was found to be less stable based on DFT.  

\newpage
\bibliography{sample}